\documentclass[fleqn,usenatbib]{mnras}
\usepackage[T1]{fontenc}
\DeclareRobustCommand{\VAN}[3]{#2}
\let\VANthebibliography\thebibliography
\def\thebibliography{\DeclareRobustCommand{\VAN}[3]{##3}\VANthebibliography}

\usepackage{graphicx}	
\usepackage{amsmath}	
\usepackage{amssymb}	
\usepackage{siunitx,fancyvrb,xcolor,bm,upgreek,tikz,fontawesome}
\DeclareSIUnit\parsec{pc}

\newcommand{\ud}{\mathrm{d}}
\newcommand{\ue}{\mathrm{e}}

\usepackage{xcolor}

\definecolor{lime}{HTML}{A6CE39}
\DeclareRobustCommand{\orcidicon}{\hspace{-3mm}
	\begin{tikzpicture}
	\draw[lime, fill=lime] (0,0) 
	circle [radius=0.16] 
	node[white] {\hspace{0.1mm}{\fontfamily{qag}\selectfont \tiny ID}};
	\draw[white, fill=white] (-0.07,0.1) 
	circle [radius=0.01];
	\end{tikzpicture}
	\hspace{-5mm}
}

\foreach \x in {A, ..., Z}{\expandafter\xdef\csname orcid\x\endcsname{\noexpand\href{https://orcid.org/\csname orcidauthor\x\endcsname}
		{\noexpand\orcidicon}}
}

\title[Radio photons from PBHs]{Background of radio photons from primordial black holes}
\author[S. Mittal and G. Kulkarni]{
Shikhar Mittal\ \orcidA{}\ \,\thanks{E-mail: shikhar.mittal@tifr.res.in}
and Girish Kulkarni\ \orcidB{}\ \,\thanks{E-mail: kulkarni@theory.tifr.res.in}
\\
Tata Institute of Fundamental Research, Homi Bhabha Road, Mumbai 400005, India
}

\date{Accepted 2022 January 03. Received 2022 January 03; in original form 2021 October 22}

\pubyear{2021}

\begin{document}
\label{firstpage}
\pagerange{\pageref{firstpage}--\pageref{lastpage}}
\maketitle
\begin{abstract}
We compute the isotropic radiation background due to Hawking emission from primordial black holes (PBHs), and examine if this background is a viable option in explaining the excess radiowave background observed by the ARCADE2 and LWA1 experiments at $\lesssim \SI{1}{\giga\hertz}$. We find that even under the extreme assumption that all of the dark matter is in the form of PBHs, the radio brightness temperature induced by Hawking evaporation of PBHs is $\mathcal{O}(10^{-46})\,$K, highly subdominant compared to the cosmic microwave background. The main reason for this is that for PBHs in the mass range $\sim\num{e12}$--$\SI{e14}{\kilo\gram}$, which can be constrained by Hawking emission, the spectrum peaks at $10^7$ to $\SI{e5}{\electronvolt}$. As the Hawking spectrum is power law suppressed towards lower energies, negligible flux of $\si{\micro\electronvolt}$ photons is obtained. The peak of the Hawking spectrum shifts to lower energies for higher masses, but the number density is low and so is the specific intensity. Because Hawking emission from PBHs is thus unable to explain the observed excess radio background, we also consider the alternative possibility of radio emission from gas accretion onto supermassive PBHs. These PBHs can readily produce strong radio emission that could easily explain the ARCADE2/LWA1 excess.
\end{abstract}

\section{Introduction}
The anomalous 21-cm signal observed by the Experiment to Detect the Global Epoch of reionization Signal (EDGES) collaboration requires a strong Lyman-$\alpha$ background \citep{Mittal} along with either excess cooling of the intergalactic medium \citep[see, e.g.,][]{B18} or an excess radio background \citep[see, e.g.,][]{Feng_2018}. In our previous work \citep{Mittal2021} we incorporated the excess radio background (ERB) -- above the standard cosmic microwave background (CMB) -- first observed by Absolute Radiometer for Cosmology, Astrophysics and Diffuse Emission \citep[ARCADE2,][]{Fixsen_2011} and later confirmed by Long Wavelength Array \citep[LWA1,][]{Dowell_2018}. While the origin of this background is not yet understood \citep{Singal_2018}, several candidates that could produce an ERB have been proposed, such as accreting astrophysical black holes \citep{Ewall, Ewall2}, bright luminous galaxies \citep{Mirocha_19}, annihilating axion-like dark matter particles \citep{FRASER2018, MOROI, choi} or dark photons \citep{Pospelov}, superconducting cosmic strings \citep{Brandenberger_2019, roxane}, supernova explosion of Population III stars \citep{Jana}, radiative decay of relic neutrinos to sterile neutrinos \citep{CHIANESE201964} and thermal emission from quark nugget dark matter \citep{LAWSON}.

In this work we explore the possibility of whether the ERB could have originated from light non-rotating and uncharged evaporating primordial black holes (PBHs). For this assessment we calculate the specific intensity of very low energy (frequencies $\sim\SI{1}{\giga\hertz}$) photons from PBHs of representative mass $\sim\num{e14}$ and $\SI{e25}{\kilo\gram}$. The former case is interesting as PBHs of mass $\sim\SI{e14}{\kilo\gram}$ can actually be constrained as dark matter candidate via non-observation of their Hawking-evaporated products \citep[e.g.,][]{capanema}, by observation from upcoming X-ray, gamma-ray, gravitational-wave experiments \citep{Ray:2021mxu,Wang:2020uvi,Calabrese:2021zfq,DeRomeri:2021xgy, ghosh2021} or their effect on the thermal properties of the intergalactic medium \citep{CMBevap,Stocker:2018avm,Acharya:2020jbv,Chan:2020zry, Cang_2021,Laha:2020vhg,Dutta:2020lqc}. The latter mass is not constrained by evaporation but via lensing effects such as those reported by Subaru Hyper Suprime-Cam \citep[HSC,][]{HSC1,HSC2}, Kepler satellite \citep{Kepler1, Kepler2}, Optical Gravitational Lensing Experiment \citep[OGLE,][]{ogle}, Exp\'erience de Recherche d'Objets Sombres \citep[EROS,][]{eros}, Massive Compact Halo Object \citep[MACHO,][]{macho}, Icarus \citep{icarus} and type Ia supernovae \citep{sne}. It is an interesting case to study because its Hawking emission peaks near the energy of a 21-cm photon.

This paper is organized as follows. In section~\ref{Theo} we present our computation of the specific intensity due to evaporating and accreting PBHs. We then discuss our results in section~\ref{RD}, and conclude in \ref{Con}. Our cosmological parameters are $\Omega_{\mathrm{m}}= 0.315$, $\Omega_{\mathrm{b}}=0.049$, $\Omega_\Lambda = 0.685$, $h=0.674$ and $T_{\mathrm{cmb}}=\SI{2.725}{\kelvin}$ \citep{Fixsen_2009, Planck}, where $T_{\mathrm{cmb}}$ is the CMB temperature measured today.

\section{PBH radio backgrounds}\label{Theo}
The basic equation required to calculate the isotropic background radiation is
\begin{equation}
J(E)=\frac{c}{4\pi}u(E)\,,\label{eq1}
\end{equation}
where $c$ is the speed of light, $u$ is specific energy density (energy per unit volume per unit energy) and $J$ is the solid-angle-averaged specific intensity (energy per unit area per unit time per unit energy per unit solid angle). Specific energy density measured at a time $t$ is obtained by adding all the energy emitted previously since a time $t_0$. Thus,
\begin{equation}
u[E(t)]=\int_{t_0}^{t} \epsilon[E(t')]\,\ud t'\,,\label{eq2}
\end{equation}
where $\epsilon$ is the comoving emissivity (energy per unit comoving volume per unit time per unit energy). Note that we have explicitly shown time dependence on $E$ as the energy collected by time $t$ from earlier times may be redshifted. Also, in writing equation~\eqref{eq2} we have assumed that photons propagate freely for $t>t_0$. A reasonable choice for $t_0$ would be $\sim4\times10^5$\,years (redshift $z_0\sim 1000$), which corresponds to last scattering of the CMB \citep{Arbey2019}. In order to correctly account for attenuation one must have the factor $\ue^{-\tau(t')}$ with the emissivity, where $\tau$ is the optical depth \citep{Ballesteros2019}. For low energy photons and because the Universe is mostly neutral after $z_0\sim 1000$, the main contribution to $\tau$ would be due to absorption/emission by the hyperfine states \citep{Field, MMR}. However, since this 21-cm optical depth \citep{Barkana_2005} comes out to be quite small (typically $\tau_{21\mathrm{cm}}\lesssim0.01$), we can safely assume $\ue^{-\tau(t')}\approx1$. Also note that not accounting for a finite optical depth gives us an `upper bound' on the radio background.

We will apply our formalism first to evaporating PBHs and then to accreting PBHs.
\subsection{Evaporating PBHs}
Assuming that PBHs are uniformly distributed in space the emissivity can be written as
\begin{equation}
\epsilon_{\mathrm{eva}}(E)=\int \mathcal{N}(M)\cdot E\cdot F_M(E)\,\ud M\,,\label{eq3}
\end{equation}
where the instantaneous spectrum of photons (only primary) from Hawking radiation is (in units of energy inverse and time inverse) \citep{Hawking, Page:1976df, Page:1976ki, Gibbon_1990, Gibbon_1991, MacGibbon:2007yq}
\begin{equation}
F_M(E)=\left(\frac{\ud\dot{N}_{\upgamma}}{\ud E}\right)_{M}=\frac{1}{h_\mathrm{P}}\frac{\Gamma_\upgamma}{\ue^{\beta E}-1}\,,\label{flux}
\end{equation}
where $h_{\mathrm{P}}$ is Planck's constant, $\beta=(k_{\mathrm{B}}T)^{-1}$ and $\Gamma_\upgamma$ is the greybody factor for photons. The temperature of a black hole of mass $M$ is
\begin{equation}
T=\frac{h_{\mathrm{P}} c^3}{16\pi^2 k_{\mathrm{B}}G_{\mathrm{N}} M} \equiv 1.06 \left(\frac{\SI{e10}{\kilo\gram}}{M}\right) \si{\giga\electronvolt}\,,\label{tbh}
\end{equation}
where $G_{\mathrm{N}}$ is Newton's constant and $k_{\mathrm{B}}$ is Boltzmann constant. We obtain $F_M$ from publicly available \texttt{C} code \texttt{BlackHawk}\footnote{\url{https://blackhawk.hepforge.org/}} \citep{blackhawk}. The comoving number density of black holes of masses between $M$ and $M+\ud M$ is $\mathcal{N}\,\ud M$, which we discuss later.

Let us calculate $J$ as a function of energy $E$ that will be measured today $(z=0)$. Using equations~\eqref{eq1}, \eqref{eq2} and \eqref{eq3} we get \citep{Carr10, Ballesteros2019, Arbey2019}
\begin{multline}
J_{\mathrm{eva}}(E)=\frac{c}{4\pi}\int_0^{z_{0}}\int \mathcal{N}(M)\cdot E\cdot(1+z')\cdot F_M[E\cdot (1+z')]\\\times\ud M\left|\frac{\ud t'}{\ud z'}\right|\ud z'\,,\label{eq4}
\end{multline}
where $z_\mathrm{0}$ is the redshift corresponding to the epoch beyond which the photons are expected to stream freely. We choose $z_0\sim1000$ as discussed above. For a measurement at any other epoch, one could just replace the two $(1+z')$ factors by $(1+z')/(1+z)$ and have the lower limit as $z$ instead of 0 in the $z$ integral. In writing equation~\eqref{eq4} we have made another simplifying assumption that the decrement in mass of black holes due to evaporation can be neglected (good enough for BHs of mass greater than $\SI{e12}{\kilo\gram}$).

We first write a general formula of $J$, i.e., for an extended mass distribution of black holes \citep{Carr17}. Let the mass function be denoted by $\ud n/\ud M$, which has the dimensions of volume inverse and mass inverse. It can be normalised under the `extreme' assumption that all of the dark matter is in the form of PBHs, so that
\begin{equation}
A\int_{M_{\mathrm{min}}}^{M_{\mathrm{max}}}M\frac{\ud n}{\ud M}\ud M=\rho_{\mathrm{dm}}\,,
\end{equation}
where $\rho_{\mathrm{dm}}$ is the dark matter density today and $A$ is the normalisation constant. We can also define a probability function as
\begin{equation}
\psi(M)=\frac{A}{\rho_{\mathrm{dm}}}M\frac{\ud n}{\ud M}\,,\label{eq5}
\end{equation}
where $\psi(M)\,\ud M$ can be interpreted as the probability for a black hole to have a mass between $M$ and $M+\ud M$. For the special case of a monochromatic distribution (MCD), i.e., no spread in the masses or in other words all the black holes are of the same mass $M_0$, we have
\begin{equation}
\psi(M)=\delta(M-M_0)\,.
\end{equation}

We now write the normalised mass function (in units of volume inverse and mass inverse) of black holes required in equation~\eqref{eq4} as
\begin{equation}
\mathcal{N}(M)=\frac{\rho_{\mathrm{dm}}}{M}\psi(M)\,.\label{eq6}
\end{equation}
Using equations~\eqref{eq4} and \eqref{eq6} we finally get the generalised expression for the specific intensity of primary photons, propagating freely since the time of emission from the evaporating PBHs, measured today
\begin{multline}
J_{\mathrm{eva}}(E)=\frac{c}{4\pi}\rho_{\mathrm{dm}}E\int_0^{z_0}\int_{M_{\mathrm{min}}}^{M_{\mathrm{max}}} \frac{1}{M}F_M[E(1+z')]\\\times\psi(M)\,\ud M\frac{\ud z'}{H(z')}\,,\label{eq7}
\end{multline}
where $H(z)$ is the Hubble function. For low energies, we can use Rayleigh--Jeans limit to also define a radio brightness temperature corresponding to the above specific intensity. Thus,
\begin{equation}
T_{\mathrm{b}}(E)\equiv\frac{h_\mathrm{P}^3c^2}{2k_{\mathrm{B}}}\frac{J(E)}{E^2}\,,\label{bt}
\end{equation}
which is applicable when $J$ is expressed in units of energy per unit time per unit area per unit solid angle per unit energy. This temperature may be used as an excess background in order to enhance the 21-cm signal.

\subsubsection{Analytical Estimate for Low-Mass PBHs}\label{2.1}
Here we develop the formalism for the special case of ultralight PBHs, such as of mass $M=\SI{e14}{\kilo\gram}$. For very low energies the greybody factor for photons is \citep{Gibbon_1990}
\begin{equation}
\Gamma_{\upgamma}\sim\frac{1024\pi^4}{3}\left(\frac{G_{\mathrm{N}}ME}{h_{\mathrm{P}}c^3}\right)^4\,.\label{gbf}
\end{equation}
The number flux, equation~\eqref{flux}, for low energy limit becomes
\begin{equation}
F_M(E)\approx \frac{1}{h_{\mathrm{P}}}\frac{1024\pi^4}{3}\left(\frac{G_{\mathrm{N}}ME}{h_{\mathrm{P}}c^3}\right)^4\left(\frac{k_{\mathrm{B}}T}{E}\right)\,,\label{eq17}
\end{equation}
where we used $(1+\beta E)$ for $\ue^{\beta E}$. For a $\SI{e14}{\kilo\gram}$ PBH the peak energy is $\sim\SI{100}{\kilo\electronvolt}$ (see equation~\ref{tbh}) whereas we are interested in energies $\sim \si{\micro\electronvolt}$. Thus, we are in the regime of $E\ll k_{\mathrm{B}}T$ where the exponential can be Taylor expanded to first order.

For MCD of black holes of mass $M$, equation~\eqref{eq7} reduces to
\begin{equation}
J_{\mathrm{eva}}(E)=\frac{c}{4\pi}\rho_{\mathrm{dm}}\frac{E}{M}\int_0^{z_0}F_M[E(1+z')]\frac{\ud z'}{H(z')}\,.\label{j}
\end{equation}
Using equations~\eqref{tbh}, \eqref{bt}, \eqref{eq17} and \eqref{j} we get
\begin{equation}
T_{\mathrm{b}}(E)=\frac{8\pi}{3}\frac{G_{\mathrm{N}}^3M^2E^2}{k_{\mathrm{B}}h_{\mathrm{P}}c^6}\rho_{\mathrm{dm}}\int_0^{z_0}\frac{(1+z')^3}{H(z')}\ud z'\,.
\end{equation}
Putting $\rho_{\mathrm{dm}}=3H_0^2\Omega_{\mathrm{dm}}/(8\pi G_{\mathrm{N}})$, where $\Omega_{\mathrm{dm}}=\Omega_{\mathrm{m}}-\Omega_{\mathrm{b}}$, and assuming a matter dominated universe for $z'\leqslant z_{0}$ we get
\begin{equation}
T_{\mathrm{b}}(E)\approx\frac{2}{5}\frac{G_{\mathrm{N}}^2M^2E^2}{k_{\mathrm{B}}h_{\mathrm{P}}c^6}\frac{\Omega_{\mathrm{dm}}}{\sqrt{\Omega_{\mathrm{m}}}}H_0(1+z_{0})^{5/2}\,.\label{analitical}
\end{equation}

\subsection{Accreting PBHs}
Accretion of matter onto supermassive black holes is thought to power the active galactic nuclei (AGN). The non-thermal luminosity of these objects span a broad range of frequency bands including radio. Several mechanisms have been proposed to explain the radio output, such as synchrotron emission by relativistic jets \citep{begelman, Panessa2019}. Observations imply that for optically thin regime, sources for synchrotron emission give rise to power law form for specific intensity with spectral index $\sim-0.6$, i.e., $J(\nu)\propto \nu^{-0.6}$ \citep{ishi}. In terms of brightness temperature this is equivalent to saying $T_{\mathrm{b}}\propto\nu^{-2.6}$, which is same as the index for ERB reported by ARCADE2/LWA1 \citep{Fixsen_2011, Dowell_2018}. This makes radio emitting supermassive black holes well-motivated candidates for our purpose. It has been briefly discussed for PBHs by \citet{Hasinger_2020}. In works by \citet{Ewall} radio emission was considered from accreting astrophysical black holes.

We construct the comoving radio emissivity as comoving number density of PBHs times specific luminosity (luminosity per unit frequency), i.e.,
\begin{equation}
\epsilon_{\mathrm{acc}}(\nu)=nl_{\mathrm{R}}(\nu)\,,    
\end{equation}
The specific luminosity can be estimated by the fundamental plane of black hole activity which connects specific radio luminosity to luminosity in X-ray band and black hole mass \citep[e.g.][]{merloni, Wang_2006}. The fundamental plane relation is calibrated at low redshifts, but given the observed diversity in AGN SEDs and it represents a conservative assumption for the high redshifts that we study here.

If $L_{\mathrm{X}}$ represents the total X-ray luminosity in $0.1$--$\SI{2.4}{\kilo\electronvolt}$ band then \citep{Wang_2006}
\begin{equation}
\log_{10}\left.\left(\frac{\nu l_{\mathrm{R}}}{L_{\mathrm{E}}}\right)\right|_{\nu=\SI{1.4}{\giga\hertz}}=0.86\log_{10}\left(\frac{L_{\mathrm{X}}}{L_{\mathrm{E}}}\right)-5.08\,,\label{fp}
\end{equation}
where $L_{\mathrm{E}}=1.26\times10^{31} M/\mathrm{M}_{\odot}$W is the Eddington luminosity. Given that the synchrotron radio emission follows a power law with index $-0.6$, the general specific radio luminosity in the vicinity of $\nu=\SI{1.4}{\giga\hertz}$ can be written as
\begin{equation}
l_{\mathrm{R}}(\nu)=\left(\frac{\nu}{\SI{1.4}{\giga\hertz}}\right)^{-0.6}l_{\mathrm{R}}(\nu=\SI{1.4}{\giga\hertz})\,.
\end{equation}
Without going into the details of accretion mechanism which gives rise to luminosity, we simply write $L_{\mathrm{X}}$ as $f_{\mathrm{X}}\lambda L_{\mathrm{E}}$, where $\lambda$ is the Eddington ratio (ratio of bolometric to Eddington luminosity) and $f_{\mathrm{X}}$ is the ratio of $L_{\mathrm{X}}$ to bolometric luminosity.

Putting everything together we get the comoving radio emissivity due to accreting PBHs as
\begin{multline}
\epsilon_{\mathrm{acc}}(E)=\num{5.65e19}f_{\mathrm{duty}}(f_{\mathrm{X}}\lambda)^{0.86}\left(\frac{f_{\mathrm{pbh}}\rho_{\mathrm{dm}}}{\SI{1}{\kilo\gram\metre^{-3}}}\right)\\\times\left(\frac{E}{\SI{5.79}{\micro\electronvolt}}\right)^{-0.6}\si{\second^{-1}\metre^{-3}}\,,\label{rem}
\end{multline}
where we converted specific luminosities from per unit frequency basis to per unit energy basis and used $f_{\mathrm{pbh}}\rho_{\mathrm{dm}}/M$ for the number density of PBHs of mass $M$. We have also inserted a duty cycle, which is the probability that a black hole is active at a particular time. We suppress the emissivity's $z$ dependence as we do not account for any explicit redshift dependence on the right hand side.

The emissivity in equation~\eqref{rem} appears to be independent of PBH mass but the dependence is actually encoded in the Eddington ratio, the duty cycle \citep[e.g.,][]{Shankar_2008, Rai09}, and the $f_{\mathrm{pbh}}$. In this work we take $f_{\mathrm{duty}}=10^{-2}$, $\lambda=0.1$ (typical of supermassive black holes) \citep{Shankar_2008} and $f_{\mathrm{X}}=0.1$. PBHs in mass range $10^5$--$10^{12}\,\mathrm{M}_{\odot}$ are constrained by dynamical effects. We adopt the strongest limit for, say a $10^{8}\,\mathrm{M}_{\odot}$ PBH, which is about $f_{\mathrm{pbh}}\sim 10^{-4}$ \citep{Carr_1999, Carr18, florian}.

The specific intensity due to the emissivity given in equation~\eqref{rem} is \citep{Ewall}
\begin{equation}
J_{\mathrm{acc}}(E,z)=\frac{c}{4\pi}(1+z)^3\int_z^{z_0}\frac{\epsilon_{\mathrm{acc}}(E')}{1+z'}\,\frac{\ud z'}{H(z')}\,,\label{jacc}
\end{equation}
where $E'=E(1+z')/(1+z)$ and $z_0=1000$ as discussed previously. Since we are interested in observations made today, we will put $z=0$ for our results.

\begin{figure*}
\centering
\includegraphics[width=1\textwidth]{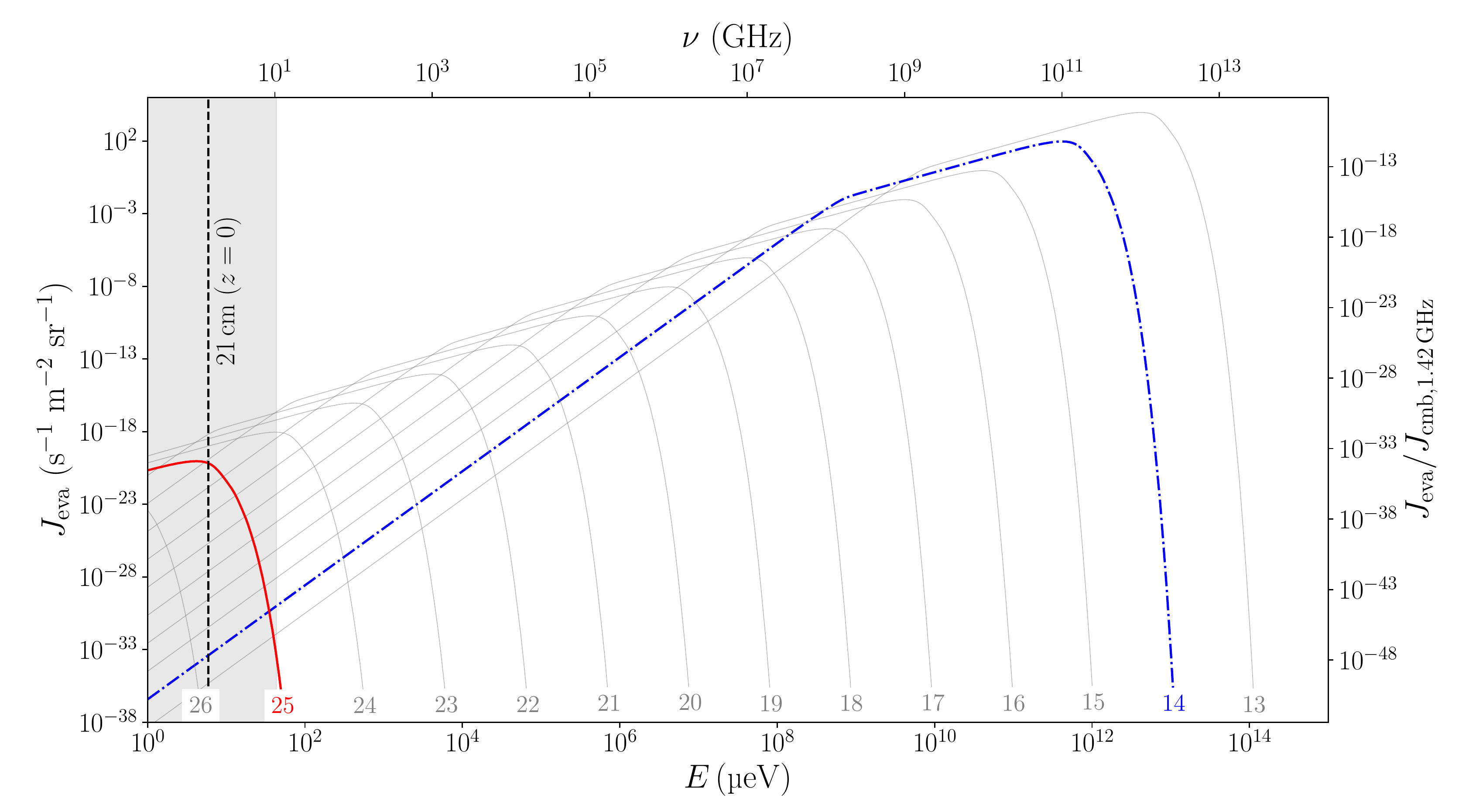}
\caption{Specific intensity of a background produced by a uniform distribution of $\num{e14}$ (dash-dotted blue) and $\SI{e25}{\kilo\gram}$ (solid red) PBHs. The grey shaded region roughly covers the frequency range over which an ERB has been seen \citep{Dowell_2018}. For illustration we have compared the specific intensity with that of CMB at $E_{21\mathrm{cm}}=\SI{5.9}{\micro\electronvolt}$ which is $J_{\mathrm{cmb}}(E_{21\mathrm{cm}})=\SI{2.53e12}{\second^{-1}\metre^{-2}\steradian^{-1}}$. For lower masses the dash-dotted blue curve would shift rightwards. Similarly for heavier black holes the solid red curve would go leftwards. In both the cases $J_{\mathrm{eva}}(E_{21\mathrm{cm}})$ would be even smaller. For comparison we have shown the results for various other masses by grey lines. The labels show $\log_{10}(M/\mathrm{kg})$.}\label{1e14}
\end{figure*}

\section{Results and Discussion}\label{RD}
We discuss three cases: primordial black holes of masses $\SI{e14}{\kilo\gram}$, $\SI{e25}{\kilo\gram}$ and supermassive of the order of $10^8\,\mathrm{M}_{\odot}$. We will always assume that the distribution is monochromatic. First two masses are for evaporating case and the last is for accreting case. For the demonstration purpose we will evaluate our results at the energy corresponding to wavelength (frequency) $\SI{21}{\centi\metre}$ ($\SI{1.42}{\giga\hertz}$), $E_{21\mathrm{cm}}=\SI{5.9}{\micro\electronvolt}$. 

\subsection{PBHs of mass \texorpdfstring{$\SI{e14}{\kilo\gram}$}{text}}
Let us consider a MCD of black holes of mass $\SI{e14}{\kilo\gram}$, for which, as mentioned earlier, the photon spectrum peaks at $\sim\SI{100}{\kilo\electronvolt}$, which falls towards far right of the energy we are interested in. The Hawking spectrum is power law suppressed on the left of its peak so we expect very small numbers. The calculation discussed in section~\ref{2.1} is applicable for this mass. Using equation~\eqref{analitical} we get 
\begin{equation}
T_{\mathrm{b}}(E_{21\mathrm{cm}})=\SI{0.78e-46}{\kelvin}\,.    
\end{equation}
Note that this estimate agrees reasonably with a more sophisticated numerical calculation using \texttt{BlackHawk}, which gives $T_{\mathrm{b}}(E_{21\mathrm{cm}})=\SI{4.25e-46}{\kelvin}$. In terms of specific intensity we get
\begin{equation}
J_{\mathrm{eva}}(E_{21\mathrm{cm}})=\SI{4.01e-34}{\second^{-1}\metre^{-2}\steradian^{-1}}\,.    
\end{equation}
Compare this with specific intensity of CMB, which is just given by a blackbody form,
\begin{equation}
J_{\mathrm{cmb}}(E)=\frac{2}{h_\mathrm{P}^3c^2}\frac{E^3}{\ue^{E/k_{\mathrm{B}}T_{\mathrm{cmb}}}-1}\,.
\end{equation}
At $E=E_{21\mathrm{cm}}$ and $z=0$ we get
\begin{equation}
J_{\mathrm{cmb}}(E_{21\mathrm{cm}})=\SI{2.53e12}{\second^{-1}\metre^{-2}\steradian^{-1}}\,.    
\end{equation}
Figure~\ref{1e14} depicts this graphically. For the lower masses considered in our previous work \citep{Mittal2021}, specific intensity or the brightness temperature will be even smaller as evident from equation~\eqref{analitical}.

\subsection{PBHs of mass \texorpdfstring{$\SI{e25}{\kilo\gram}$}{text}}
Usually masses in range $\sim\num{e12}$--$\SI{e14}{\kilo\gram}$ are constrained via their evaporated products. For higher masses it becomes irrelevant since mass loss rate is inversely proportional to mass squared. However, heavier PBHs maybe an interesting case because for some mass the primary photon emission could peak at $E_{21\mathrm{cm}}$. In general, for black hole temperature $T$, peak occurs at $5.77k_{\mathrm{B}}T$ \citep{MacGibbon:2007yq} which can be used to estimate the corresponding mass as follows
\begin{equation}
5.77\times1.06\left(\frac{\SI{e10}{\kilo\gram}}{M}\right)=\num{5.9e-15}\,,
\end{equation}
giving $M\approx\SI{e25}{\kilo\gram}$ or $\num{5e-6}\,\mathrm{M}_{\odot}$. The strongest constraint on this PBH seems to stem from microlensing measurements using Subaru/HSC \citep{HSC1, HSC2}, which sets the limit to $f_{\mathrm{pbh}}\lesssim0.1$. However, we will still assume this to be 1 as in previous case, in order to consider an extreme-case scenario.

Because we are near the peak, the greybody factor estimate in equation~\eqref{gbf}, and hence equation~\eqref{analitical}, is not applicable. We take the help of \texttt{BlackHawk} and find that for this mass we get
\begin{equation}
J_{\mathrm{eva}}(E_{21\mathrm{cm}})=\SI{6.74e-21}{\second^{-1}\metre^{-2}\steradian^{-1}}\,,    
\end{equation}
and the corresponding brightness temperature is
\begin{equation}
T_{\mathrm{b}}(E_{21\mathrm{cm}})=\SI{7.18e-33}{\kelvin}\,.    
\end{equation}
This is roughly 13 orders of magnitude higher than the result for $\SI{e14}{\kilo\gram}$ PBH, but it still is negligible in comparison to CMB. Even though the peak is near $E_{21\mathrm{cm}}$, the number of such heavy PBHs is greatly reduced. We again see that we do not get an appreciable intensity. In figure~\ref{1e14} we show our results for mass $\SI{e25}{\kilo\gram}$ by the solid red line. If we considered a mass higher than $\SI{e25}{\kilo\gram}$, the peak would be leftwards of $E=E_{21\mathrm{cm}}$ and the specific intensity would be exponentially suppressed (see equation~\ref{flux}).

We see that the primary emission from non-rotating monochromatic distributed PBHs of any mass do not produce an appreciable low energy photon background. Rotating PBHs evaporate more strongly compared to non-rotating ones \citep{Chandra, Taylor} but still not enough to cover 30 to 40 orders of magnitude. We can also consider extended mass distribution of PBHs, such as that motivated by \citet{Press} form for dark matter haloes which is based on the spherical gravitational collapse in density perturbations \citep{Young_2020, sureda21}. As an example, when the mass function is normalised to $\rho_{\mathrm{dm}}$ between masses $M_{\mathrm{min}}=\SI{e12}{\kilo\gram}$ and $M_{\mathrm{max}}=\SI{9e16}{\kilo\gram}$ -- a range in which Press--Schechter from resembles a power law -- we get $T_{\mathrm{b}}(E_{21\mathrm{cm}})\sim \SI{e-31}{\kelvin}$. Thus, even for broad mass distributions like Press--Schechter form, we end up with the same conclusion as for MCD. Also note that in our results until now we assumed $f_{\mathrm{pbh}}=1$, but with stronger limits our background will be even smaller, precisely by a factor of $f_{\mathrm{pbh}}$.

An important point to note is that for low energies the secondary spectra for Hawking emission may become important due to hadronisation of primary particles \citep{Gibbon_1990, Gibbon_1991, Carr10, comptel}. However, we have not taken it into account. The secondary photons result either because of annihilation of oppositely charged particles (chiefly e$^\pm$, $\upmu^{\pm}$, $\uppi^{\pm}$) or decay of unstable primary particles ($\upmu^{\pm}$, $\uppi^{0,\pm}$). Recently, the code \texttt{BlackHawk} was updated to calculate the secondary spectra to energies as low as $\sim\SI{1}{\kilo\electronvolt}$ \citep{Arbey_2021}, which still is 9 orders of magnitude higher than the energy of our interest. Calculation of the secondary spectra is beyond the scope of this work and we leave it for future study.

\subsection{Supermassive PBHs}
\begin{figure}
\centering
\includegraphics[width=1\linewidth]{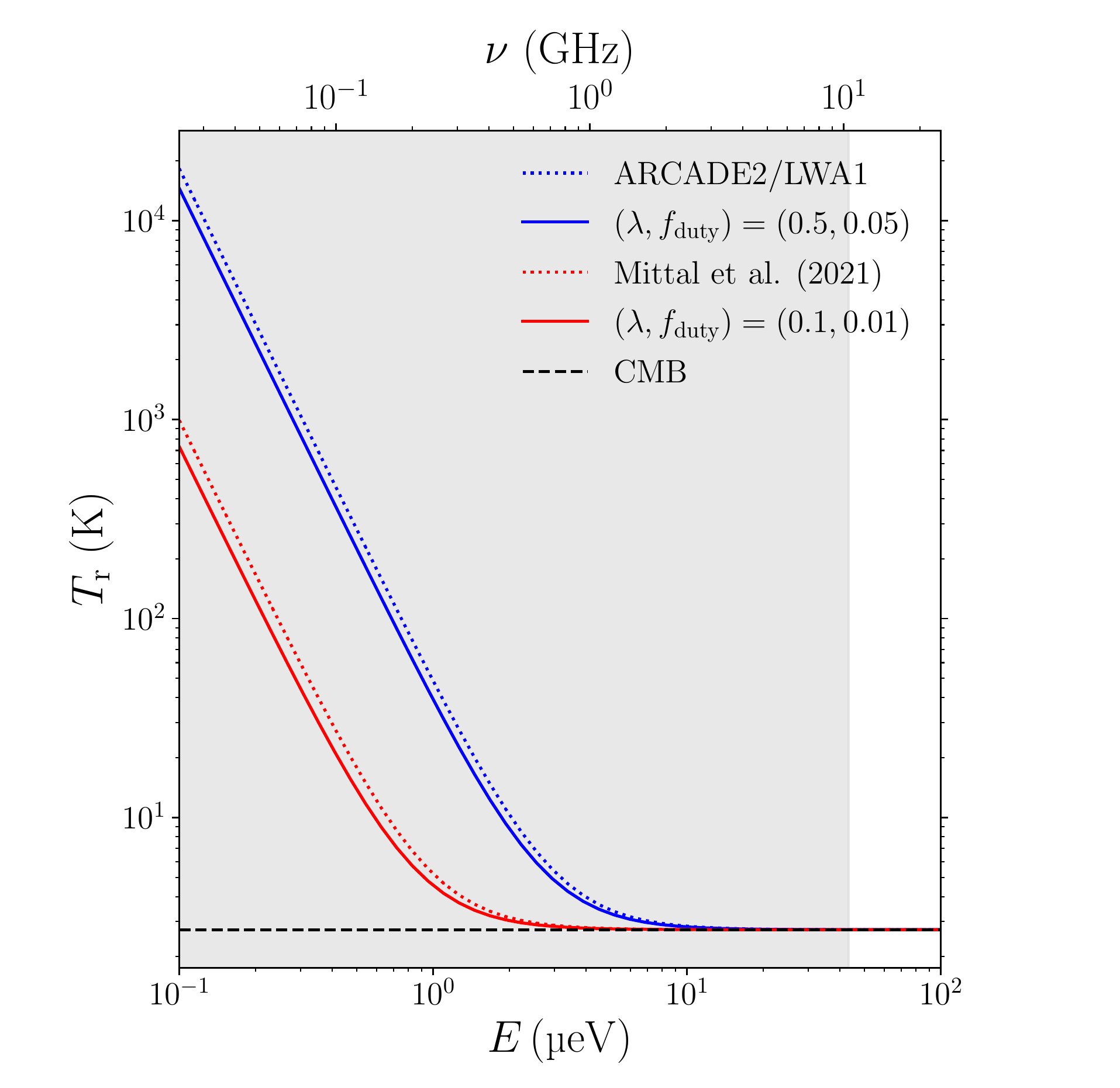}
\caption{The net background temperature ($T_\mathrm{r}=T_{\mathrm{b}}
+T_{\mathrm{cmb}}$) generated by radio emission due to accretion onto supermassive PBHs. For $\lambda=0.1, f_{\mathrm{X}}=0.1, f_{\mathrm{duty}}=10^{-2}$ and $f_{\mathrm{pbh}}=10^{-4}$ (solid red) we can easily explain the ERB required (dotted red) to explain the amplitude of the EDGES measurement of the global cosmological 21-cm signal. However, the actual excess observed which is $\sim 20$ times larger (dotted blue) can also be explained with $\lambda=0.5$ and $f_{\mathrm{duty}}=5\times10^{-2}$ (other two parameters being the same). The CMB temperature is shown in dashed black for reference. The grey shaded region roughly covers the frequency range over which an ERB has been seen \citep{Dowell_2018}.}\label{acc}
\end{figure}

It is clear that as far as primary Hawking emission is concerned, PBHs play no role at radio wavelengths. However, accretion onto heavier PBHs as dark matter candidate may contribute to radio emission as we will see here. Applying equation~\eqref{jacc} at $E=E_{21\mathrm{cm}}$ and $z=0$ we get
\begin{equation}
J_{\mathrm{acc}}(E_{21\mathrm{cm}})=\SI{1.7e10}{\metre^{-2}\second^{-1}\steradian^{-1}}\,,
\end{equation}
which can be scaled to other energies by the factor $(E/E_{21\mathrm{cm}})^{-0.6}$. Using equation~\eqref{bt} we get the radio brightness temperature due to accreting PBHs to be
\begin{equation}
T_{\mathrm{b}}(E_{21\mathrm{cm}})\approx\SI{0.02}{\kelvin}\,.
\end{equation}
Compare the above result (which is 5 per cent) with the upper limit set by LWA1, which when extrapolated to $E=E_{21\mathrm{cm}}$ gives $\sim\SI{0.5}{\kelvin}$. It is easy to see that with a slight readjustments in the empirical factors we can even explain the full LWA1 ERB (shown by dotted blue line in figure~\ref{acc}). Quantitatively, we would need to increase the product $f_{\mathrm{duty}}(f_{\mathrm{X}}\lambda)^{0.86}f_{\mathrm{pbh}}$ by a factor of 20, which can be done using $f_{\mathrm{duty}}=5\times10^{-2}$ and $\lambda=0.5$ (result shown by solid blue). Note that the fundamental plane relation (equation~\ref{fp}) used here was calibrated for radio-quiet AGN by \citet{Wang_2006}. The relation is more robust and reliable compared to that for the radio-loud sample. However, if we use the relation as such it will only change numbers but not our conclusions. Alternatively, the radio-loud emission can also be estimated as follows. Radio-loud AGN \citep{Keller} constitute $\sim 10$ per cent of the total AGN population \citep{Banados_2015} and are roughly 1000 times brighter than radio-quiet in relevant wavelengths. This would give us an overall factor of 100 for the radio background, showing that AGNs can produce strong radio backgrounds.

In figure~\ref{acc} we show the net background temperature, i.e., $T_{\mathrm{r}}=T_{\mathrm{b}}
+T_{\mathrm{cmb}}$ at $z=0$ as a function of photon energy for our chosen parameters by the solid red line. We compare this with the 5 per cent -- as inferred by \citet{Mittal2021} to explain the EDGES result when X-ray heating is included -- of the LWA1 limit shown by the dotted red line. Also shown for reference is the CMB temperature by the dashed black line.

We have shown that radio emission due to accretion can explain the LWA1 ERB and hence possibly the EDGES 21-cm signal \citep[however, see][]{Sharma2018}. Also, note that Previously, accreting PBHs have been constrained by their heating effect on baryons \citep{hektor, mena, Yang2021, yang20212, Pablo21}. However, a more consistent analysis would include both, heating and background enhancement for the 21-cm signal \citep[such as][although for astrophysical BHs]{Ewall2}. Given the observational data it may be possible to obtain proper constraints on the properties of accreting PBHs.

\section{Conclusion}\label{Con}
We studied the specific intensity and the corresponding radio brightness temperature of primary photons from evaporating primordial black holes (PBHs), with representative masses of $\num{e14}$ and $\SI{e25}{\kilo\gram}$. The smaller of these values is the highest mass that can be constrained by its Hawking evaporated products, while the higher value is of interest because its Hawking spectrum peaks at the energy corresponding to a 21-cm photon. The brightness temperature values today for these masses in a monochromatic distribution assuming all dark matter is composed of PBHs ($f_{\mathrm{pbh}}=1$) are $\sim\SI{e-46}{\kelvin}$ and $\sim\SI{e-33}{\kelvin}$, respectively, at an energy of $E_{21\mathrm{cm}}=\SI{5.9}{\micro\electronvolt}$. In both the cases this is extremely small compared to cosmic microwave background temperature measured today. Our main conclusion is that primary photons from evaporating PBHs cannot explain the excess radio background such as that observed by LWA1. An alternative scenario in which radiation is produced due to gas accretion on supermassive PBHs does potentially work, however, and could readily explain the observed excess. We have made the codes used in this work publicly available\footnote{\url{https://github.com/shikharmittal04/Radio-photons-PBHs.git}}.

\section*{Acknowledgment}
We thank the anonymous referee for suggestions that helped improve the paper. We acknowledge stimulating discussions with J\'er\'emy Auffinger, Basudeb Dasgupta, Nishita Desai, Avery Meiksin and Anupam Ray. GK gratefully acknowledges support by the Max Planck Society via a partner group grant. GK is also partly supported by the Department of Atomic Energy (Government of India) research project with Project Identification Number RTI 4002.

\section*{Data availability}
No new data were generated or analysed in support of this research.

\bibliographystyle{mnras}
\bibliography{Biblo}

\bsp
\label{lastpage}
\end{document}